\documentclass[10pt,conference]{IEEEtran}
\IEEEoverridecommandlockouts
\usepackage{tikz}
\usepackage{amsmath}

\usepackage{filecontents}
\usepackage{booktabs}
\usepackage{graphicx}
\usepackage{units}
\usepackage{blindtext}
\usepackage{xcolor,xspace}
\usepackage{amsmath}
\usepackage{paralist}
\usepackage{mdframed}
\usepackage{pict2e}
\usepackage{mdframed}
\usepackage{listings}
\usepackage{xcolor}
\usepackage{threeparttable}
\usepackage{array}
\usepackage{pifont}
\usepackage{balance}
\usepackage{verbatim}
\usepackage{array}
\usepackage{booktabs}
\usepackage{float}
\usepackage{listings}
\usepackage{subfigure}

\def\BibTeX{{\rm B\kern-.05em{\sc i\kern-.025em b}\kern-.08em
    T\kern-.1667em\lower.7ex\hbox{E}\kern-.125emX}}

\usepackage{nccmath}
\usepackage{soul}
\usepackage{multirow}
\usepackage{graphicx}
\usepackage{array}
\definecolor{littlegreen}{RGB}{193,236,193}
\definecolor{littlegrey}{RGB}{128,128,128}
\definecolor{altgreen}{RGB}{97,166,46}

\usetikzlibrary{arrows}
\usetikzlibrary{shapes}
\newcommand*\circled[3]{\tikz[baseline=(char.base)]{
    \node[scale=.7,shape=circle,draw,inner sep=1pt,fill=#2,minimum size=.1cm,text=#3] (char) {#1};}}
    
    \usepackage{color,colortbl}
    
    \newcommand{\ignore}[1]{}

    \newcommand{\acro}{ISC-FLAT\xspace}

    \newcommand{\request}{{\ensuremath{\sf{\mathcal Req}}}\xspace}
    \newcommand{\report}{{\ensuremath{\sf{\mathcal R}}}\xspace}
    \newcommand{\app}{{\ensuremath{\sf{\mathcal App}}}\xspace}

    \newcommand{\prv}{{\ensuremath{\sf{\mathcal Prv}}}\xspace}
    \newcommand{\vrf}{{\ensuremath{\sf{\mathcal Vrf}}}\xspace}
    \newcommand{\RA}{{\ensuremath{\sf{\mathcal RA}}}\xspace}
    \newcommand{\SAU}{{\ensuremath{\sf{ SAU}}}\xspace}

    \newcommand{\adv}{{\ensuremath{\sf{\mathcal Adv}}}\xspace}
    \newcommand{\chl}{{\ensuremath{\sf{\mathcal Chl}}}\xspace}

\newcommand{\CFlog}{{\ensuremath{\sf{\mathcal CF_{Log}}}}\xspace}

\newcommand{\CFA}{CFA\xspace}
\newcommand{\CFG}{CFG\xspace}
\newcommand{\tee}{TEE\xspace}
\newcommand{\ISM}{ISM\xspace}

\definecolor{codegreen}{rgb}{0,0.6,0}
\definecolor{codegray}{rgb}{0.5,0.5,0.5}
\definecolor{codepurple}{rgb}{0.58,0,0.82}
\definecolor{backcolour}{rgb}{0.95,0.95,0.92}

\lstdefinestyle{mystyle}{
    commentstyle=\color{codegreen},
    keywordstyle=\color{blue},
    numberstyle=\tiny\color{codegray},
    stringstyle=\color{codepurple},
    basicstyle=\fontsize{9}{9}\ttfamily,
    frame=single,
    breakatwhitespace=false,         
    breaklines=true,                 
    captionpos=b,                    
    keepspaces=true,                 
    numbers=none,                    
    numbersep=4pt,                  
    showspaces=false,                
    showstringspaces=false,
    showtabs=false,                  
    tabsize=2       
    }
    
    \lstdefinestyle{customasm}{
        belowcaptionskip=1\baselineskip,
        frame=single, 
        frameround=tttt,
        xleftmargin=\parindent,
        language=[x86masm]Assembler,
        basicstyle=\footnotesize\ttfamily,
        commentstyle=\itshape\color{green!60!black},
        tabsize=4,
        numbers=left,
        numbersep=8pt,
        stepnumber=1,
        numberstyle=\tiny\color{gray}, 
        columns = fullflexible,
        }

\begin{document}

\title{ISC-FLAT: On the Conflict Between Control Flow Attestation and Real-Time Operations\\
}

\author{\IEEEauthorblockN{Antonio Joia Neto}
\IEEEauthorblockA{
\textit{Rochester Institute of Technology}\\
aj4775@rit.edu}
\and
\IEEEauthorblockN{Ivan De Oliveira Nunes}
\IEEEauthorblockA{
\textit{Rochester Institute of Technology}\\
ivanoliv@mail.rit.edu}
}

\maketitle




\begin{abstract}
  
    The wide adoption of IoT gadgets and Cyber-Physical Systems (CPS) makes embedded devices increasingly important. While some of these devices perform mission-critical tasks, they are usually implemented using Micro-Controller Units (MCUs) that lack security mechanisms on par with those available to general-purpose computers, making them more susceptible to remote exploits that could corrupt their software integrity. Motivated by this problem, prior work has proposed techniques to remotely assess the trustworthiness of embedded MCU software. Among them, Control Flow Attestation (\CFA) enables remote detection of runtime abuses that illegally modify the program's control flow during execution (e.g., control flow hijacking and code reuse attacks).
    
    Despite these advances, current \CFA methods share a fundamental limitation: they preclude \textit{interrupts} during the execution of the software operation being attested. Simply put, existing \CFA techniques are insecure unless interrupts are disabled on the MCU. On the other hand, we argue that the lack of interruptability can obscure \CFA usefulness, as most embedded applications depend on interrupts to process asynchronous events in real-time.
    
    To address this limitation, we propose \underline{I}nterrupt-\underline{S}afe \underline{C}ontrol \underline{Fl}ow \underline{At}testation (\acro): a \CFA technique that is compatible with existing MCUs (i.e., does not require hardware changes) and enables interrupt handling without compromising the authenticity of \CFA reports. Similar to other CFA techniques that do not require customized hardware modifications, \acro leverages a Trusted Execution Environment (TEE) (in particular, our prototype is built on ARM TrustZone-M) to securely generate unforgeable \CFA reports without precluding applications from processing interrupts.
    We implement a fully functional \acro prototype on the ARM Cortex-M33 MCU and demonstrate that it incurs minimal runtime overhead when compared to existing TEE-based \CFA methods that do not support interrupts.
\end{abstract}


\section{Introduction}\label{sec:intro}

From IoT gadgets to vehicular safety-critical sensors, society has grown accustomed to the pervasiveness of embedded devices. Naturally, this increased reliance is accompanied by a growing risk of embedded software compromise. 
Unfortunately, prevention of software compromises in embedded devices is especially challenging because they are typically implemented using (one or several) Micro-Controller Units (MCUs). Due to strict cost and energy budgets, MCUs lack security mechanisms commonly found in higher-end general-purpose CPUs, such as strong separation of privilege levels and memory management units to support virtual memory and isolation. Furthermore, embedded devices are often deployed in multitudes, sharing the same (potentially vulnerable) software. Unsurprisingly, the insecurity of embedded software has already resulted in several attacks, including massive Distributed Denial of Service (DDoS)~\cite{antonakakis2017understanding, deogirikar2017security} and large-scale exploits with economical and life-threatening consequences~\cite{pycroft2018security,ring2015connected,stuxnet,giraldo2016integrity}.

Since preventive approaches are often too costly or unfeasible in resource-constrained embedded devices, security services that enable remote detection of software compromise have attracted attention in recent years~\cite{kuang2022survey}. Remote Attestation (\RA)~\cite{seshadri2006scuba,perito2010secure,Viper2011,smart,vrased,simple,trustlite,tytan,hydra,brasser2016remote,erasmus,smarm,ibrahim2017seed,carpent2018temporal,pistis,scraps} is one such service that enables a Verifier (\vrf) to assess the trustworthiness of the software executing on a remote low-end embedded device -- called a Prover (\prv). In its simplest form (a.k.a. ``static \RA'' or ``\RA of binaries''), \RA offers means to detect illegal modifications to the binary installed and running on \prv. However, by itself, it provides no information about the order in which instructions within the binary execute at runtime. In particular, control flow hijacks~\cite{abadi2009control} and code reuse attacks~\cite{shacham2007geometry,jit_code_reuse} (e.g., via return-oriented programming~\cite{rop,rop_no_ret}) can  change the order in which instructions execute without modifying the binary. As the binary remains the same, such attacks cannot be detected by static \RA.

To address this limitation, Control Flow Attestation (\CFA)~\cite{cflat,lofat,litehax,atrium,oat,tiny-cfa,scarr,recfa} augments static \RA to provide \vrf with an unforgeable ``control flow proof'', containing the order in which the instructions of the attested binary have executed. As such, \CFA enables the detection of control flow hijacks and code reuse attacks, even when these attacks do not modify the installed binary.

\CFA defines a Control Flow Graph (CFG). Nodes in the CFG are sequences of non-branching instructions. Edges represent control flow transfers (e.g., jumps, calls, returns, etc.). \CFA techniques work by tracking the path taken in the program's CFG during execution. This model assumes that all instructions within the same CFG node are guaranteed to execute sequentially. However, when interrupts are enabled, this assumption is falsifiable because interrupts can cause control flow transfers asynchronously within any given node in the CFG. As a result, existing \CFA techniques either assume~\cite{cflat,oat} or enforce~\cite{tiny-cfa,dialed} disablement of all interrupts in the MCU.  On the other hand, most real-time applications are interrupt-based~\cite{vahid2001embedded}. This conflict can make current \CFA methods impractical in many settings.

\textbf{Contributions.} In this work, we aim to reconcile \CFA with the real-time needs of MCU application domains. To that end, we propose \acro: an \underline{I}nterrupt-\underline{S}afe \underline{C}ontrol \underline{Fl}ow \underline{At}testation method. \acro leverages ARM TrustZone-M to isolate side-effects of external interrupts from the execution being attested. Since TrustZone-M is available in several ARM CPUs, \acro is readily deployable on ``off-the-shelf'' MCUs without requiring additional customized hardware. To the best of our knowledge, \acro is the first TEE-based \CFA approach to securely support interrupts while maintaining \CFA integrity. In sum, the anticipated contributions of this paper are three-fold:

\begin{compactitem}
  \item We formulate and characterize the conflict between real-time applications and existing TEE-based \CFA methods. To motivate the need for interrupt-safe \CFA, we demonstrate interrupt-based attacks on current \CFA designs.
  \item We propose \acro, a TEE-based \CFA design that supports interrupts without compromising the integrity of underlying \CFA proofs. At its core, \acro implements a TrustZone-based secure interrupt dispatcher that leverages TrustZone's Nested Vectored Interrupt Controller (NVIC) to interpose itself between any interrupt trigger and its respective service routine. The dispatcher saves the necessary context of the interrupted task within TrustZone's Secure World and verifies that this context is resumed appropriately when the service routine ends. This design ensures \CFA integrity for the interrupted task without making service routines part of \acro Trusted Computing Base (TCB).
  \item We implement and evaluate \acro on an ``off-the-shelf'' ARM Cortex-M33 MCU, equipped with TrustZone-M. Our experimental results demonstrate \acro\ {\it quasi}-negligible overhead when compared to an existing \CFA technique that does not support secure interrupts.
  To foster future research in this topic, we make \acro implementation publicly available at~\cite{repo}.
\end{compactitem}

\section{Background \& Related Work}  

\subsection{ARM TrustZone-M} \label{sec: preliminaires: Arm TrustZone}  

ARM TrustZone is a Trusted Execution Environment (TEE) included in modern ARM CPUs. It partitions the System on Chip (SoC) hardware and software into Secure and Non-secure regions (called ``Worlds''). Resources belonging to the Secure World are isolated from the Normal (Non-Secure) World, resulting in a secure environment for executing security-critical functions and storing sensitive data. TrustZone's hardware controllers prevent the Normal World from accessing physical memory regions assigned to the Secure World.

TrustZone-M defines the security state of memory segments (i.e., whether a segment in the address space belongs to the Secure or Normal World) by using a combination of the Secure Attribution Unit (\SAU), and the Implementation Defined Attribution Unit (IDAU) to enforce spatial isolation. IDAU is a fixed memory map defined by the manufacturer, while  \SAU is programmable by the Secure World.

A number of prior efforts aim to leverage TrustZone-M to enhance embedded system security from various perspectives, including low latency secure interrupts~\cite{sbis}, real-time system availability guarantees~\cite{wang2022rt}, Address-Space Layout Randomization (ALSR) without requiring memory management units~\cite{fASLR}, and support for virtualization~\cite{pinto2019virtualization}.
For a comprehensive overview of TrustZone's architecture, we refer the reader to~\cite{pinto2019demystifying}.

\subsection{Interrupts \& TrustZone-M NVIC}\label{sec: preliminaires:  Interruptions TZ}

TrustZone-M capable MCUs process all interrupts using two separated Interrupt Vector Tables (IVTs) for the Secure and Normal Worlds. They are managed by an integrated controller called  Nested Vector Interrupt Controller (NVIC). Each interrupt can be assigned as Secure or Non-secure by setting a register named Interrupt Target Non-secure (NVIC\_ITNS), which is only programmable in the Secure World. In addition, the IVTs can share the same priority level, or secure interrupts can have priority over non-secure ones. The interrupt pipeline follows the standard execution flow if an interrupt is triggered while the CPU is in the same security state as the interrupt. If a Secure interrupt is triggered while the CPU is in the non-secure state, the CPU ignores the Non-Secure IVT and redirects execution to the Interrupt Service Routine (ISR) address stored in the Secure IVT, while automatically pushing the registers of the interrupted task to the non-secure stack. Recent related work~\cite{sbis,oliveira2021utango} takes advantage of the NVIC controllers. Specifically, the NVIC\_ITNS register is used to enforce specific interrupt states as a requirement for the additional security features.   

\subsection{Static Remote Attestation (\RA)} \label{sec: preliminaires:  Binary Attestation}  

Static \RA (a.k.a. ``\RA of binaries'') allows a verifier (\vrf) to determine the integrity of an application's binary running on an untrusted remote platform (\prv), i.e., \vrf is able to detect illegal modifications to the binary. \RA is also a building block for other services, including Control Flow Attestation (\CFA), Data Flow Attestation ($DFA$), and Proof of Execution ($PoX$)~\cite{apex}. For instance, since TEE-based \CFA requires binary instrumentation (see details in Section~\ref{sec:C-flat}), \RA is necessary as a part of \CFA to guarantee that the instrumentation has not been maliciously removed.

\begin{figure}[h]      
    \centering
    \vspace{-1em}
    \includegraphics[width=0.6\columnwidth]{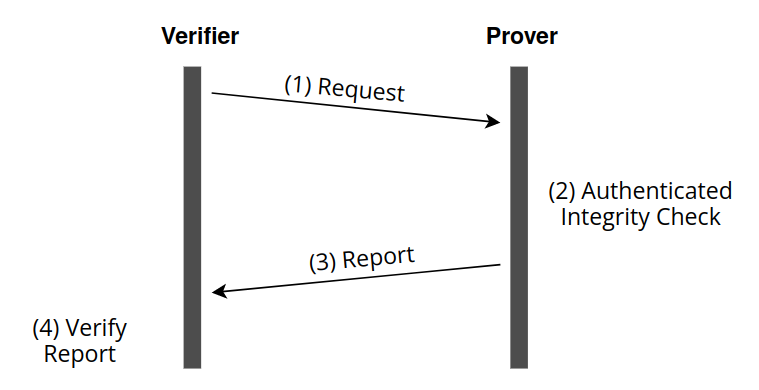}
    \vspace{-1em}
    \caption{\RA interaction}
    \vspace{-0.6em}
\label{fig:RA}   
\end{figure}

As depicted in Figure~\ref{fig:RA}, \RA is a challenge-response protocol wherein \vrf sends an attestation request to \prv containing a cryptographic challenge ($Chl$) (e.g., a random nonce) to guarantee that the \prv will generate a unique, timely response to this request. \prv receives $Chl$ and computes an authenticated integrity check (e.g., a MAC or a signature) over its own program memory and $Chl$, and sends the resulting report back to \vrf. Finally, \vrf checks the received report against the expected value (i.e., the expected binary for \prv).

Static \RA architectures are typically classified as software-based, hardware-based, or hybrid. Software-based \RA~\cite{seshadri2004swatt,choi2007proactive,VIEIRASTEINER201943} does not require hardware support. Instead, it leverages precise timing, along with strong assumptions about direct communication and adversarial silence, to detect malware on \prv. To relax these assumptions, hardware-based attestation relies on a hardware root of trust for measurement and reporting to produce unforgeable attestation reports securely. This category includes TPM-based approaches~\cite{xu2011remote}, Intel SGX \cite{wang2017enabling,kucab2021remote}, and ARM TrustZone \cite{ahn2020design}. Hardware-based designs provide a high level of security. However, they are restricted to devices that support these custom features. Between hardware and software-based, hybrid \RA~\cite{perito2010secure,trustlite,smart,tytan,vrased,de2021toctou} requires minimal hardware trust anchors to provide strong security guarantees on resource-constrained devices. Besides just attesting the integrity of a single device's binary, collective \RA methods focus on groups of devices  \cite{carpent2017lightweight,sap,ambrosin2018pads,asokan2015seda}. Carpent et al. \cite{carpent2018temporal,smarm} introduce attestation for self-relocating malware.

\subsection{Control Flow Attestation (\CFA)} \label{sec:C-flat}  

\CFA augments \RA by including a proof of the sequence in which instructions were executed in the attested binary. Consequently, \CFA can detect runtime attacks that hijack the program's control flow but do not modify \prv's code (e.g., the well-known return-oriented programming and code-reuse attacks~\cite{eternal-war}). \CFA is achieved by tracking and logging the control flow path taken during the execution of the attested binary. While \CFA does not actively protect the system against control flow attacks, it provides \vrf with information to detect any control flow attack during the execution. As such, \vrf can accurately decide if a result produced by this execution (e.g., a sensed value) is trustworthy.

In \CFA, \vrf requests from \prv an authenticated proof that: {\bf (i)} \prv indeed executed the expected binary (denoted \app) in a timely manner, i.e., after the most recent request from \vrf; and {\bf (ii)} there were no control flow attacks during \app execution. Optionally, the proof may also include any execution results (e.g., a sensed value) produced by \app execution on \prv, allowing \vrf to assess the trustworthiness of this result based on the \CFA verification. We refer to this ``proof'' as the \CFA report.
\textbf{We define a \CFA report as \textit{unreliable} if the control flow path taken during \app execution differs from the one contained in the \CFA report.} Therefore, an unreliable report could make an attack execution oblivious to \vrf.

\begin{figure}[h]      
  \centering
  \includegraphics[width=.95\columnwidth]{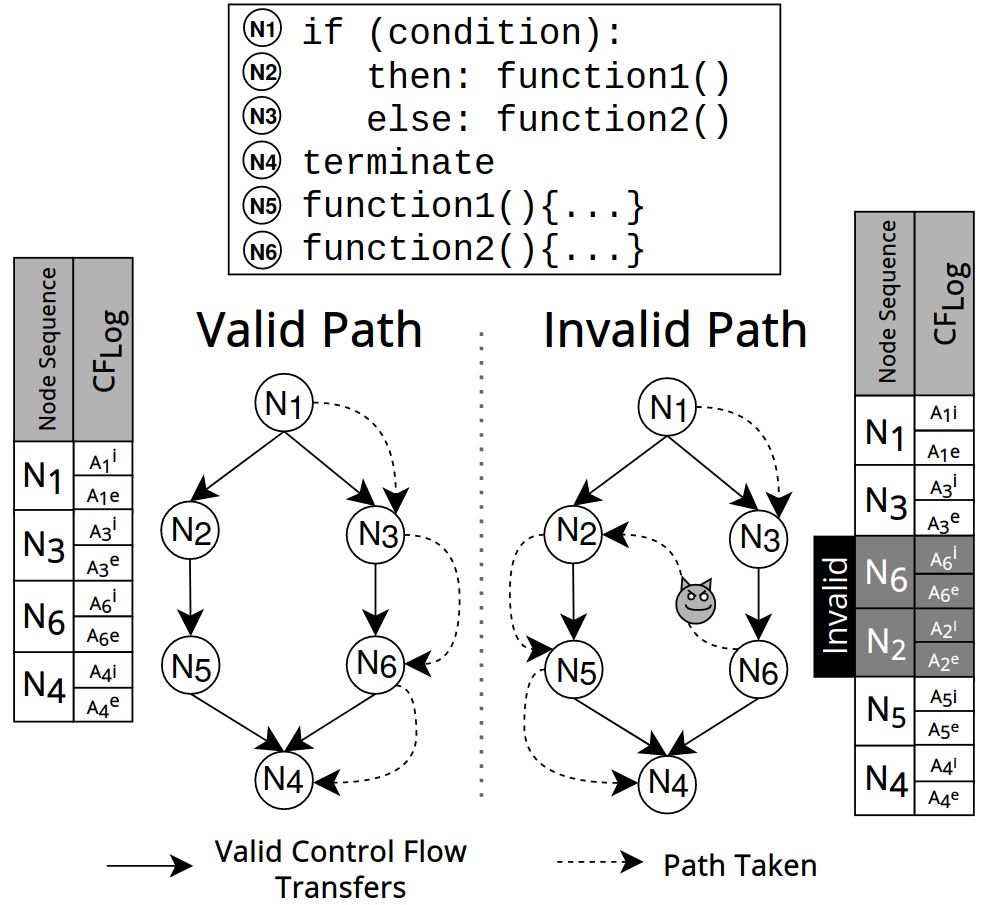}
  \vspace{-1em}
  \caption{The figure illustrates an example of a valid (left) and an invalid (right) control flow execution with their respective generated \CFlog.}       
\label{fig:cfa_example}
\vspace{-1em}
\end{figure}

TEE-based \CFA methods~\cite{cflat,oat,liu2019log} use automated binary instrumentation to build an authenticated control flow log (\CFlog) containing all control flow transfers, i.e., the destination address of all the branching instructions (e.g, \texttt{jumps}, \texttt{calls}, \texttt{returns}) taken during execution. \vrf can then check \CFlog to decide if the control flow transfers were valid. Every node in the attested program's CFG is instrumented with additional trampoline instructions that trap the execution into the Secure World. Within the Secure World, the branch destination is appended to \CFlog. \CFlog itself is stored in the Secure World protected memory. This approach assumes that only branching instructions can modify the control flow. Therefore, sequences of regular (non-branching) instructions are treated as ``blocks''. A block followed by a branching instruction defines a node of the CFG (a block can be empty in the case of two sequential branching instructions). When interrupts are disabled, this approach works well because instructions within a block are unable to modify the program's control flow. However, when interrupts are enabled, any number of control flow transfers may occur within the execution of a single block. Even worse: these transfers would not be saved to \CFlog and therefore not noticed by \vrf. In Section~\ref{sec : problem_settings} we elaborate on this issue and further motivate its importance.

Figure \ref{fig:cfa_example} exemplifies \CFA. $N_1$ - $N_6$ are nodes in the CFG of the attested application. Solid arrows represent legal control flow transfers, and dashed arrows represent control flow paths taken during a particular program execution. When a node is executed entirely, the additional (instrumented) instructions within the node produce two new entries in \CFlog (denoted $A_x^i$ and $A_x^e$, for node $N_x$). $A_x^i$ represents the memory address of the first instruction of the node $N_x$, and $A_x^e$ represents the destination of $N_x$'s branching instruction (i.e., $N_x$ last instruction). The example in the left part of the figure shows a valid execution that traverses existing edges in the program's CFG. The path is composed by the node sequence \{$N_1,N_3,N_6,N_4$\} and generates the \CFlog entries \{$A_1^i,A_1^e,A_3^i,A_3^e,A_6^i,A_6^e,A_4^i,A_4^e$\}. By inspecting \CFlog, \vrf can check the entries sequence to validate the control flow path and the fact that each node executes in its entirety (due to in order appearance of pairs \{$A_x^i$,$A_x^e$\}, i.e., valid entry and exit points of each node). The right part of the figure exemplifies an invalid execution path, containing an illegal transition from $N_6$ to $N_2$. For instance, this could be caused by an attack (e.g., a buffer overflow) that overwrites \texttt{function2()}'s return address to point to \texttt{function1()}. In this second case,  the control flow path is \{$N_1,N_3,N_6,N_2,N_5,N_4$\}, generating the \CFlog entries \{$A_1^i,A_1^e,A_3^i,A_3^e,A_6^i,A_6^e,A_2^i,A_2^e,A_5^i,A_5^e,A_4^i,A_4^e$\}. By inspecting \CFlog, \vrf can identify the control flow abuse due to the invalid sequence of entries \{$A_6^i,A_6^e,A_2^i,A_2^e$\} present in \CFlog.

C-FLAT \cite{cflat} is the earliest work to introduce \CFA and propose an instrumentation-based method using TrustZone as a TEE. OAT \cite{oat} optimizes \CFA to reduce the number of measured control flow transfers and enforce data integrity. ReCFA \cite{recfa} proposes to condense the control flow events to generate a compressed \CFA report. LO-FAT~\cite{lofat}, LiteHAX~\cite{litehax}, and Atrium\cite{atrium} propose to attest the control flow without instrumenting the executable binary, by adding customized hardware modules, such as a branch filter, a loop monitor, a hash controller, and a hash lookup. Tiny-CFA \cite{tiny-cfa} proposes a hybrid design for low-end MCUs that uses instrumentation and only requires the formally verified hardware from support from APEX proofs of execution~\cite{apex}.

All of these approaches assume that attested applications execute atomically, meaning that they cannot be interrupted. To deal with interrupts, OAT~\cite{oat} suggests instrumenting all ISRs along with the attested application. While this simplifies the evaluation of the \CFlog, it does not provide information on whether the interrupts have resumed correctly. Moreover, it is not suitable for MCUs running multiple applications. To the best of our knowledge, our work presents the first secure \CFA architecture that can be applied to interrupt-prone and real-time systems and applications.

\section{Problem Statement: Interrupts \& \CFA} \label{sec : problem_settings}      

This section discusses why current TEE-based \CFA methods generate {\bf unreliable reports} when interrupts are enabled. It also demonstrates associated attacks in practice. As discussed in Section~\ref{sec:intro}, this is an important problem because disabling all interrupts on the embedded device is often impractical.

\subsection{Interrupt-based Attack Examples}

We now present interrupt-based \CFA attacks and discuss the fundamental limitations explored by these attacks.
To demonstrate their practicality, we have also implemented an open-source example of the most general attack case (i.e., example 3 below). For more details on this implementation, see our repository at \cite{repoattack}.

\begin{figure}[thp!]           
  \centering     
  \vspace{-1em}
  \includegraphics[width=\columnwidth]{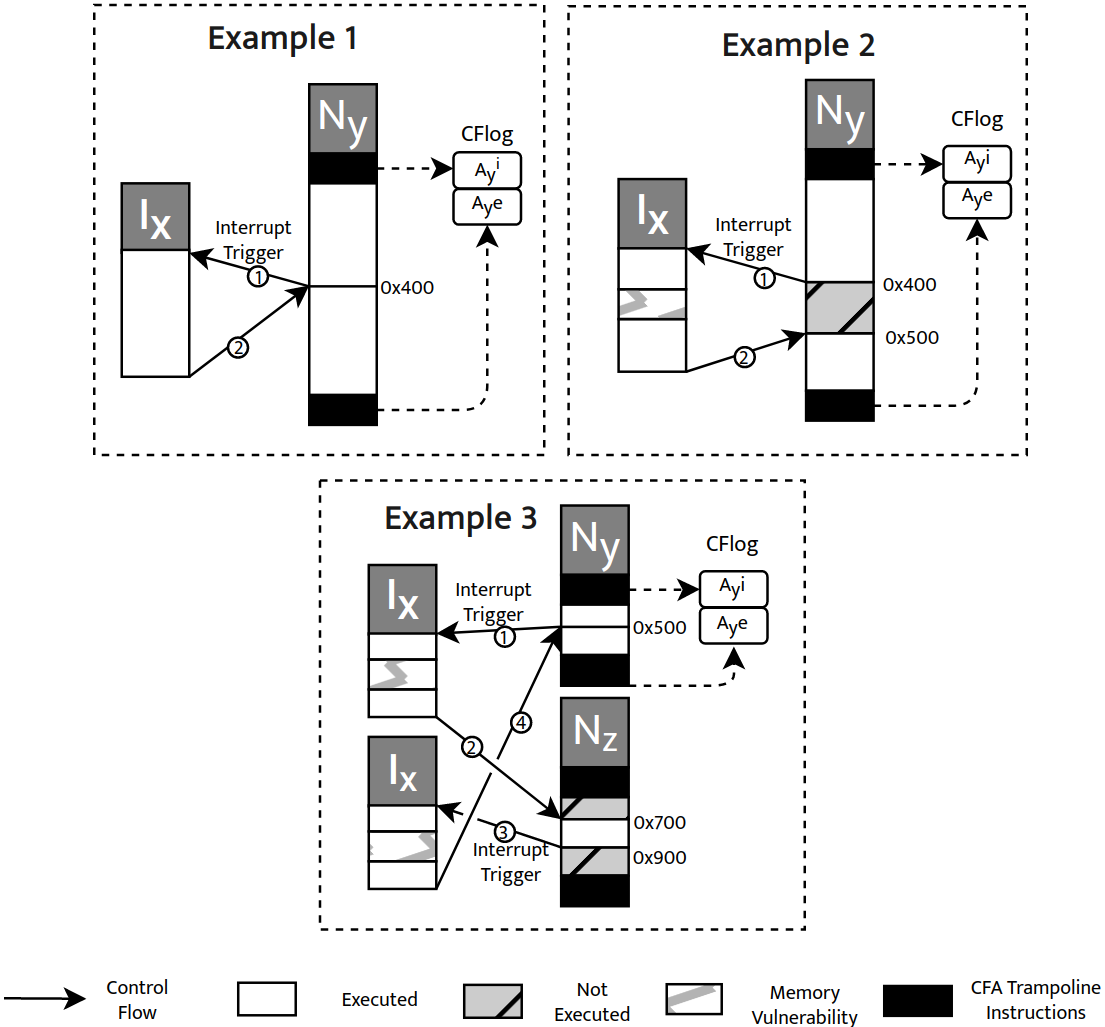}
  \vspace{-2em}
  \caption{Interrupt-based attacks on \CFA.}
  \vspace{-0.8em}
  \label{fig:problem_setting_examples}    
\end{figure}

Figure~\ref{fig:problem_setting_examples} illustrates three execution possibilities for \prv when interrupts are enabled. $I_x$ represents the implementation of an untrusted ISR unrelated to the attested application \app. Note that \vrf is only concerned with and knowledgeable about \app's binary and is oblivious to low-level system ISRs, such as $I_x$. Consequently, $I_x$ is not instrumented, and its control flow transfers are not appended to \app's \CFlog (because \vrf cannot interpret them). $N_y$ and $N_z$ represent nodes (lists of sequential instructions) within \app. Whenever they execute, new entries are added to \CFlog. \app is instrumented to generate two new \CFlog entries per node: one before executing the node's first instruction and one before the branching instruction (i.e., the last instruction) of the node. This instrumentation is in line with the TEE-based \CFA methods discussed in Section~\ref{sec:C-flat}. Without interrupts, once a node's execution starts, instructions within the node run sequentially, and this approach generate a reliable \CFlog. However, this can not be guaranteed when interrupts are enabled, as discussed below.

Example 1 in Figure~\ref{fig:problem_setting_examples} illustrates a benign interrupt control flow transfer, where $I_x$ does not tamper with \app control flow path. Execution starts from the first instruction in $N_y$, which is a trampoline to the Secure World to add new entry $A_y^i$ to \CFlog. Before reaching the instruction in address 0x400, the interrupt is triggered, and execution is redirected to $I_x$. After $I_x$ execution, $N_y$ resumes correctly from address 0x400 and proceeds sequentially, finally reaching the second trampoline instruction that adds entry $A_y^e$ to \CFlog. After executing the Node $N_y$, \CFlog will have two new log entries \{$A_y^i$,$A_y^e$\}. Without interrupts or with a benign $I_x$ that resumes $N_y$ correctly, these two new \CFlog entries indicate that all instructions within $N_y$ executed.

We now consider the case where $I_x$ is vulnerable/malicious, as illustrated in Example 2 of Figure \ref{fig:problem_setting_examples}. In this case, Adversary (\adv) could corrupt the return address of $I_x$ (e.g., by exploiting a buffer overflow within $I_x$ code), modifying it from 0x400 to 0x500. Once $I_x$ returns, the CPU resumes the execution in address 0x500, skipping all the instructions between these addresses and consequently modifying the control flow and the behavior of \app. However, in this case, the generated \CFlog is the same as in Example 1 since the trampoline instructions of $N_y$ are still reached in the same order.

Example 3 of Figure \ref{fig:problem_setting_examples} illustrates an even worse case, where \adv leverages control over interrupt configurations. For instance, timer-based interrupts can be configured to interrupt \prv precisely after a selected number of instructions (see our implementation in \cite{repo}). In Example 3, \adv leverages this capability to change the return address of a first instance of $I_x$ to the address 0x700 located inside a different \app Node: $N_z$. In addition, the attacker configures the interrupt $I_x$ to be triggered again after a specific number of clock cycles: enough to execute the desired number of instructions but less than enough to reach the trampoline instruction that appends information to \CFlog. For instance, the second interrupt occurs after executing the instructions from 0x700 to 0x900 within $N_z$. At that point, $I_x$ execution is triggered again. This time, the attacker changes $I_x$ return address to 0x400 (the original return address of the first interrupt). At the end of Node $N_y$, the new \CFlog entries generated by the described process would be exactly the same as Example 1. Therefore, \vrf would consider this report legal. In reality, however, this attack strategy allows \adv to execute any number of sub-sequences of instructions (gadgets) within \app nodes, implying arbitrary code reuse and a complete modification of \app's intended behavior. Given an appropriate gadget set in \app (which occurs in most programs), this can result in undetected Turing-complete malicious behavior~\cite{rop}.

\subsection{What makes existing \CFA vulnerable to interrupt attacks?}

Based on the previous discussion, we elucidate the fundamental limitations that enable interrupt-based attacks. This discussion serves as a guideline to the design of \acro, presented next, in Section~\ref{sec:methodology}.

\textbf{Instrumentation without Atomicity.} TEE-based \CFA methods assume that instructions within a node are executed atomically. Therefore, the binary is only instrumented in each node's first and last instruction position. Interrupts falsify this premise, enabling control flow transfers at any point within the node. An na\"ive solution to this problem would be to instrument every instruction within the node. However, such an approach would incur extremely high overhead, requiring context switches between Secure and Normal Worlds for every executed instruction in \app.

\textbf{Untraceable ISRs.} ISRs are external to \app; thus, their behavior is not reflected in \CFlog. Since they are untraceable, security mechanisms should be in place to ensure non-interference of ISRs of \ the \app's execution.

\textbf{No Stack Isolation.} Unfortunately, ISRs run as privileged code in the Normal World (see~\cite{pintouser} for related discussion). Therefore, they have access to all of the Normal World's stack (including control flow associated data and registers pushed to the stack during function calls). Conversely, they can modify interrupt configurations and also misconfigure Memory Protection Units (MPU) when applicable. These issues imply that any protection against malicious/compromised ISRs must be enforced by \prv's Secure World.

\section{ISC-FLAT: Overview}
\label{sec:overview}

\acro architecture comprises three modules to support secure interruptable \CFA: (i) Interrupt Safety Module (\ISM), (ii) \CFA Measurement Engine, and (iii) Remote Verification Engine.

\begin{compactitem}
    \item \textbf{Interrupt Safety Module (\ISM):} This module aims to augment TEE-based \CFA with the capability to generate reliable reports while enabling interrupts. To this end, \ISM securely initializes attestation and creates a dispatcher inside the Secure World that interposes itself between interrupt triggers and the respective ISR execution. The dispatcher configures protections to the interrupted \app before the ISR can execute.
    
    \item \textbf{CFA Measurement Engine:} The measurement engine tracks \app's control flow and manages \CFlog. This is achieved through instrumentation of \app's binary. Before \app's deployment on \prv, an automated instrumentation script adds instructions to \app's assembly code for each node of \app's CFG (as described in Section~\ref{sec:C-flat}). This instrumentation invokes a Secure World component that appends new entries to \CFlog according to the control flow path taken during \app's execution.
   
    \item \textbf{Remote Verification Engine:} This module is executed by \vrf to analyze the \CFA report and detect violations either due to software control flow attacks or malicious interrupts during \app's execution.
\end{compactitem}

\begin{figure}[h]      
  \centering
  \includegraphics[width=1\columnwidth]{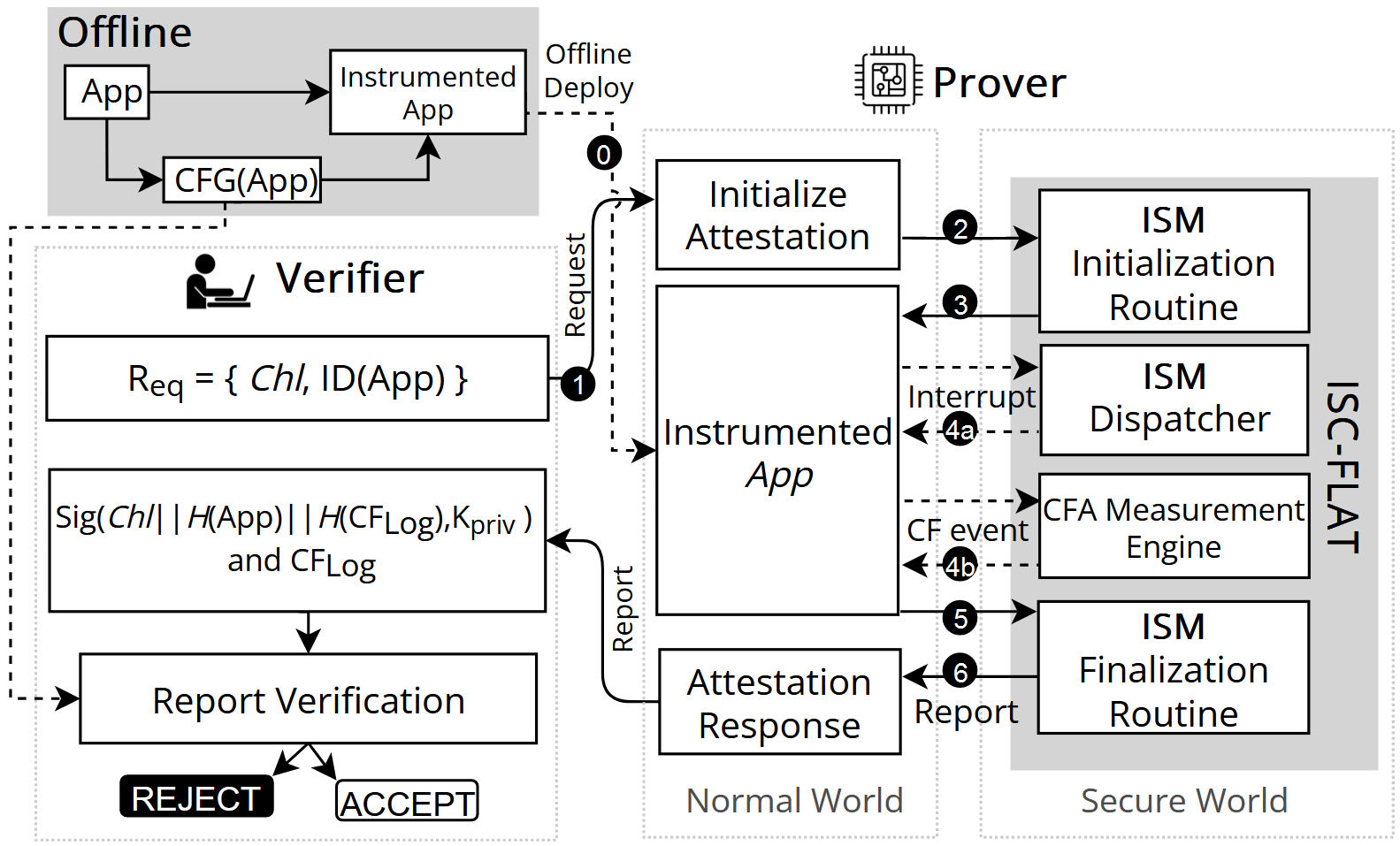}
  \vspace{-2em}
  \caption{Overview of \acro protocol.}
  \vspace{-0.6em}
\label{fig:sysmodel}   
\end{figure}

Figure \ref{fig:sysmodel} illustrates \acro's protocol.  Before being deployed on \prv, each CFG node of \app binary is instrumented with additional instructions to track the execution control flow (step \circled{0}{black}{white}). These additional instructions activate the CFA engine by adding entries to \CFlog during \app execution. 

  As indicated in step \circled{1}{black}{white}, \acro's protocol starts with \vrf locally generating a request \request : \{ \chl, $ID(\app)$ \} composed of: a challenge/nonce (\chl) and an identifier ($ID(\app)$), where \app is the application that should be executed and attested. \chl is a unique random number to provide liveness, preventing replayed \CFA reports. 
  To initiate an attested execution of \app, \vrf sends \request to \prv.

  Upon receiving \request, \prv must call the \ISM Initialization Routine, implemented in the Secure World, to create an attestation instance associated with \chl (step \circled{2}{black}{white}). Failure to do so implies the inability to produce a valid \CFA report (since the \ISM finalization routine -- detailed below -- only signs \CFA reports associated with initialized attestation instances).
  \ISM Initialization Routine also computes a hash of all the memory regions in \prv that contain \app's executable, resulting in $H(\app)$. Next, it makes the memory regions storing this binary immutable (see Section~\ref{sec:ISM} for details). Similar to \chl, $H(\app)$ is also assigned to this attestation instance. 
  \chl and $H(\app)$ are defined during initialization (before \app execution) and included in the signed \CFA report (see below).
  Therefore, any attempt to initialize an incorrect binary $\app_{\adv}$ is detectable by \vrf during verification, due to the mismatch $H(\app) \neq H(\app_{\adv})$. Finally, the \ISM Initialization Routine uses NVIC to configure all interrupts to be handled by the Secure World's \ISM dispatcher. The dispatcher, in turn, is responsible for redirecting interrupts to their real ISRs after enabling \acro's interrupt protections.

  Following \ISM Initialization Routine, \app execution starts (step \circled{3}{black}{white}). During execution, all control flow transfers are saved to \CFlog in the Secure World (step \circled{4a}{black}{white}) by the \CFA engine. Similar to \chl and $H(\app)$, \CFlog is associated to this particular attestation instance. During \app execution, due to the initial NVIC configuration, all interrupts are trapped to \ISM before their handling by their respective (untrusted) ISRs (step \circled{4b}{black}{white}). For all interrupts, \ISM Dispatcher will save \app's context and ``lock'' \app's state (including \app's stack). After this process, \ISM redirects the interrupt to its actual (untrusted) ISR, located in the Normal World. Once finalized, ISR must return to the \ISM Dispatcher module to re-enable \app's execution (i.e., unlock \app's state and stack). In this final stage, the ISM Dispatcher assures that \app's control flow and context are resumed appropriately. This protects \app's execution against (otherwise untraceable) interrupt-based \CFA attacks.

  Once \app execution is over, the \ISM Finalization Routine is 
  called (step \circled{5}{black}{white}) to produce a signed \CFA report ($\report(\app)$) containing the produced \CFlog along with \chl and $H(\app)$. Optionally, the report may include any output ($out$) produced by \app's execution. If all steps succeed, $\report(\app)$ is sent to \vrf (step \circled{6}{black}{white}). Finally, \vrf uses the Remote Verification Engine to decide on $\report(\app)$'s trustworthiness (The remote verification process is detailed in Section~\ref{sec:Analysing the report}).

\section{\acro in Detail}\label{sec:methodology}

This section further details each component in \acro.
We start by specifying the system and adversary models. Then we describe each of \acro's components, namely \CFA Measurement Engine, Interrupt Security Module (\ISM), and Remote Verification Engine.

\subsection{System Model}\label{sec:model}

We consider that \prv is a single-core, bare-metal MCU, equipped with a TEE, such as ARM TrustZone-M.
\prv hosts multiple untrusted applications, including untrusted privileged software in the form of ISRs or a simple real-time operating system (RTOS). All untrusted software modules (including the application to be attested -- \app) execute in the Normal World to keep security critical functionality (including the trusted \CFA implementation) isolated within the Secure World.
In line with the \CFA related work (see Section~\ref{sec:C-flat}), we assume the following \prv features, which are implemented by the ARM TrustZone-M (v8) architecture:

\begin{compactitem}
  \item \prv can securely store a secret key ($sk$) within the Secure World (making it inaccessible to the Normal World). The (ideally minimal) code inside the TEE's Secure World is trusted.
  \item \prv features separate IVTs for Normal and Secure Worlds. Any interrupts can be delegated to either one of the IVTs. The Secure-IVT has priority over the Normal-IVT (when an interrupt exists in both IVTs).
  \item \prv features a Non-Secure Memory Protection Unit (NS-MPU). NS-MPU is a hardware monitor that controls memory access in the Normal World. Nonetheless, the Secure World can hijack control of the NS-MPU by restricting Normal World's access to NS-MPU configuration registers\footnote{In Armv8-M is possible to restrict the Normal World from accessing the NS-MPU by marking the NS-MPU configuration memory region as belonging to the Secure World, using TrustZone-M \SAU (recall \SAU from Section~\ref{sec: preliminaires:  Interruptions TZ}).}. As discussed in Section~\ref{sec:ISM}, \acro leverages this capability to ephemerally make \app's code immutable yet executable in the Normal World.
\end{compactitem}

We assume that \vrf has knowledge of \app's binary and CFG but is unaware of other  (potentially malicious/vulnerable) application/system-level software executing on \prv. \vrf also has knowledge of the public key ($pk$) corresponding to the $sk$ (stored within \prv's Secure World).

\subsection{Adversary Model}

We consider a strong adversary (\adv) that has full control of \prv's Normal World (including the privileged mode in the Normal World). \adv can modify Normal World code (e.g., through code injection attacks) and trigger control flow hijacks and code re-use attacks.
Similarly, \adv can control interrupt configurations to call ISRs at any time~\cite{busi2020provably} and modify/corrupt ISR implementations. On the other hand, \adv is unable to tamper with Secure World-resident software and data. Conversely, it cannot disable or bypass TEE hardware-enforced access control rules and guarantees.

Compared to prior work, this threat model considers a stronger \adv that leverages interruptions and malicious Normal World code during \CFA of \app. As such, it presents a more realistic case where \CFA must securely co-exist with the real-time requirements and multiple tasks in embedded devices. We consider invasive physical attacks that modify hardware out-of-scope, as they require an orthogonal set of anti-tampering methods~\cite{ravi2004tamper}.

\subsection{CFA Measurement Engine}
\label{sec:cfa_engine}

 The measurement engine generates \CFlog and works in parallel with \ISM (described in Section~\ref{sec:ISM}) to support secure interruptable \CFA. It leverages binary instrumentation to construct \CFlog during \app execution. 

\textbf{Instrumenting Control Flow Transfers}.
Static analysis~\cite{angr} is used to generate \app's CFG, denoted $\CFG(\app)$ (recall the CFG definition from Section~\ref{sec:C-flat}).
The \CFG is used to both (i) instrument \app's binary before deployment (see below); and (ii) by \vrf to verify the \CFA reports (see Section~\ref{sec:Analysing the report}). Before deployment, each node of $\CFG(\app)$ in \app's binary is instrumented to save control flow events to \CFlog. The additional instructions are trampolines that redirect the execution to a function implemented within the Secure World that appends the control flow information to \CFlog. As \CFlog is stored within the Secure World, it cannot be directly modified by Normal World software. Specifically, the instrumentation is added before each node's first and last instructions (recall that the last instruction is the branching instruction of the node). Therefore, during execution, \CFlog will be constructed as the sequence of control flow transfers, containing two entries per node.

In each node's entry point, the instrumentation introduces a single \textit{branch and link} (\textit{bl}) instruction that branches to a Non-Secure Callable (NSC) function. The NSC function calls the \CFA engine and sends the Link Register (LR) value (containing the returning address) as an argument to be recorded in the \CFlog. The entry generated at this point represents the memory address of the node's first instruction, and its presence in \CFlog indicates that the node execution started from the first instruction.

Two cases must be considered for the instrumentation at the end of a node, depending on the nature of the node's branching instruction. If the branching instruction is a \textit{conditional branch, direct jump, or call}, the additional instruction is the same as for the entry point, and the value of the log entry represents the memory address of the node's last instruction. If the branching instruction is an \textit{indirect jump/call or return}, the entry to be logged is the (dynamically determined) branch destination. In this case, the instrumentation adds two instructions: first, it performs a \textit{push} of the destiny address to the stack and then a \textit{bl} to an NSC function that will \textit{pop} the destiny address from the stack and send it to the \CFA engine. By using this method, dynamically defined illegal branches can be detected. The instrumentation approach outlined here permits the observation of all direct and indirect branches that take place during the execution of \app. Upon completing the instrumentation process, \app can be installed on \prv, where it can operate as a Normal World application.

\textbf{Measuring the Control Flow.} Once the ISM Initialization Routine is called, it allocates a secure memory space to store \CFlog within the Secure World and sets a flag to enable the addition of new log entries. From this point until the end of \app's execution, every time the CPU reaches the trampoline instructions, it will pass the control to the Secure World, which will append the related memory addresses (entries) in the memory that comprises \CFlog.  

\begin{figure}[h]      
  \centering
  \includegraphics[width=.9\columnwidth]{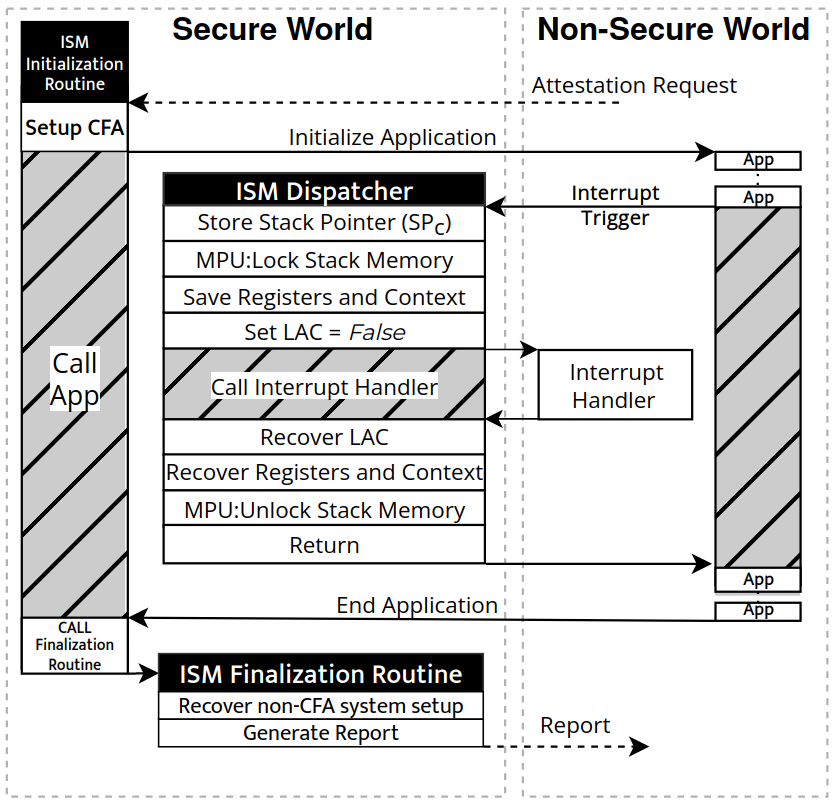}
  \vspace{-1em}
  \caption{Illustration of the \ISM workflow.}
\label{fig:isc-flat-dispatcher}
\vspace{-1em}
\end{figure}

\subsection{Interrupt Security Module (\ISM)}\label{sec:ISM}

\ISM is the key feature to make \CFA reliable when interrupts are enabled. As shown in Figure~\ref{fig:isc-flat-dispatcher}, \ISM is implemented in the Secure World to safeguard \app against all interrupts before they are handled by their untrusted ISRs in the Normal World. In doing so, \ISM assures the \CFA integrity of the interrupted \app on the following fronts: (i) protecting the stack portion containing the context and the returning address of the interrupted \app; (ii) blocking ISRs from executing/modifying code sections in \app or generating/tampering with \CFlog entries; and (iii) ensuring that \app is resumed correctly.

\ISM includes three sub-modules. \ISM Initialization Routine is responsible for all initial configurations that will guarantee the security of the attestation process and activate the secure interrupt dispatcher.
\ISM Dispatcher intermediates every ISR activation to restrict access to the memory belonging to \app as well as protecting \app control flow integrity from external interference. \ISM Finalization Routine is responsible for generating the signed \CFA report, restoring pre-attestation system configurations, and finalizing the attestation instance. 

\vspace{1mm}
\subsubsection{\bf \ISM Initialization Routine} When \prv receives the request \request from \vrf, the \ISM Initialization Routine function, located in the Secure World, must be called to initialize the attestation. This routine contemplates necessary configurations to initialize and ensure the \CFA security by following the subsequent steps:

\noindent{\bf $\bullet$ Step 1:} Check if another application is currently being attested. If so, returns an error flag.
    
\noindent{\bf $\bullet$ Step 2:} Configure TrustZone \SAU to set the memory region containing NS-MPU rule configurations as a Secure World region. This revokes Normal World permission to configure NS-MPU. 

\noindent{\bf $\bullet$ Step 3:} Configure the NS-MPU to change the permission of \app's program memory addresses to read-only. \textit{This will protect \app's binary against modifications during its execution/attestation}. 

\noindent{\bf $\bullet$ Step 4:} Hash \app's program memory, generating  $H(\app)$. \vrf will use this to later verify \app's binary integrity.
    
\noindent{\bf $\bullet$ Step 5:} Assign all interrupts to be handled by the interrupt \textit{ Dispatcher}, in the Secure World. This is achieved by setting NVIC\_ITNS. All the interrupts are set as Secure Interrupts by activating all entries in TrustZone's NVIC Secure-IVT to point to the \ISM Dispatcher. Therefore, the \ISM Dispatcher obtains the ability to interpose itself between all interrupts and the execution of their (untrusted) ISRs in the Normal World.

\noindent{\bf $\bullet$ Step 6:} Save the current value of the Normal World's Stack Pointer $SP_0$. As this is done immediately before \app execution starts, the Dispatcher can use $SP_0$ to determine the location of \app's stack and ``lock'' it, i.e., prevent modifications during the execution of eventual ISRs. 
    
\noindent{\bf $\bullet$ Step 7:} Set the Log Access Control ($LAC = True$) flag. $LAC$ indicates that the measurement engine is allowed to add new entries to \CFlog.
    
\noindent{\bf $\bullet$ Step 8:} Call \app function in the Normal World.    

\vspace{1mm}
\noindent\subsubsection{\bf \ISM Dispatcher} While \app is running, the interrupt {\it Dispatcher} handles all the interrupts before redirecting them to their original ISR implementation. The Dispatcher's goals are to: (i) track all interrupts and control their permissions to add new entries in \CFlog (ii) guarantee that the \app's context, including registers, the return address, and data in the stack, is the same after the interrupting ISR completes; (iii) prevent any attempt to bypass the Dispatcher and avoid the required context-recovery from happening. As soon as an interrupt is triggered, Secure World assumes the CPU control and creates an instance of the Dispatcher to track the interrupt. At this point, the Dispatcher saves the current Non-Secure Stack Pointer on a temporary variable ($SP_c$). Depending on the architecture, it is also necessary to store other critical registers that need protection (i.e., any registers not pushed to the stack before the interrupt). Next, the Dispatcher locks the \app's stack by setting the NS-MPU to define the region within addresses [$SP_0$, $SP_c$] as a read-only (recall that after Initialization Routine, NS-MPU cannot be modified by the Normal World). This will temporarily prevent the Normal World from overwriting the \app's stack, protecting it from ISR interference. Finally, the Dispatcher stores the previous $LAC$ value and sets $LAC = False$, implying that no entries will be appended to \CFlog until $LAC = True$ again. This prevents the ISR from adding fake entries to \CFlog.

The Dispatcher then passes control to the Normal World ISR and waits for the ISR. To find the address of the Non-Secure ISR code, the Dispatcher looks-up the NS-IVT position associated with the triggered interrupt. Note that \app execution cannot be resumed unless the ISR returns to the Dispatcher. Meanwhile, the \app's context is blocked. As a consequence, in order to produce a valid attestation report, the untrusted ISR must eventually give control back to the Dispatcher, by returning appropriately, otherwise, a valid attestation report can not be produced for \app execution. When the Dispatcher receives control back, it changes $LAC$ to its previous value and recovers all the \app's context, including the stack pointer register, to the same values as before the interrupt triggering. Finally, the previous configuration of the NS-MPU is restored, and \app execution is resumed.

\textbf{Interrupt Preemption.}  Note that a new dispatcher instance is created every time an interrupt is triggered. In the case of a preempting interrupt, multiple dispatcher instances -- each related to each active interrupt -- will exist simultaneously. Whenever the preemption is triggered, the LAC value, registers, and NS-MPU configuration belonging to the preempted task are pushed onto a stack within the Secure World. Then, the new dispatcher instance sets LAC to False and reconfigures the NS-MPU accordingly. When the preempting interruption concludes, the former NS-MPU/LAC/registers values are popped from the protected stack. 

\vspace{1mm}
\noindent\subsubsection{\bf \ISM Finalization Routine} Once the execution of \app ends, the \ISM finalization routine, implemented in the Secure World, is automatically called to generate the \CFA report and reset configurations made by the Initialization Routine at the beginning of attestation (before \app attested execution). We note that the finalization routine function is not callable by the Normal World. Thus, signing the CFA report without a prior call to initialize the attestation is impossible. The signed report is bound to the data produced and stored by the latest \ISM Initialization Routine call, including \chl and $H(\app)$. Specifically, the finalization routine performs the following actions:

 \textbf{Producing CFA Report.} The final \CFA report is denoted by $\report(\app) = \{ \sigma^{\app},\CFlog\}$. It contains all information needed to prove to \vrf which control flow path was taken by \app. \CFlog is the verbatim control flow path. The signature $\sigma^{\app}$ has the format
\begin{equation*}\small
    \sigma^{\app}=Sig^{sk}(H(\CFlog)||H(\app)||\chl),
\end{equation*}
where $Sig^{sk}$ denotes a cryptographic signature computed using Secure World's secret key $sk$. The signature is computed on $H(\CFlog)$, $H(\app)$, and \chl. Therefore, it authenticates \CFlog, \app's binary and proves the freshness of $\report(\app)$ (through \chl). Optionally, $\report(\app)$ (and respective signature) may include any results produced by \app's execution, proving that the results (e.g., sensed quantities) were produced by a trustworthy execution of \app.

\textbf{Restoring Non-CFA System Status.} The finalization routine sets $LAC = False$ to disable new additions to the \CFlog. In addition, it re-configures Non-Secure Interrupts to be handled in the Normal World by their original ISRs without going through the Dispatcher checks. The original NS-MPU  configuration also returns to its original state. 
\vspace{-1em}
\subsection{Remote Verification Engine (executed by \vrf)}
\label{sec:Analysing the report}

\textbf{Authentication}. Upon receiving $\report(\app)$ from \prv, \vrf verifies the signature using $pk$ and based on the hash of \app's expected binary, \chl, and according to the copy of \CFlog, that is included in $\report(\app)$.

\textbf{Control Flow Validation}. If all authentication checks (above) succeed, \vrf is assured that \CFlog corresponds to \app's execution on \prv. As a final step, \vrf can analyze \CFlog through a variety of means. The simplest approach is to run \CFlog through \app's legal \CFG (recall Figure~\ref{fig:cfa_example}). Furthermore, to verify backward edges (e.g., return address integrity) \vrf emulates a shadow stack~\cite{eternal-war} for \app's reported \CFlog.

\section{Security Analysis}
\label{sec:sec analysis}

\acro security argument is two-part: (1) we show security when interrupts are disabled; and (2) we show that \ISM assures that \CFlog-s produced by \acro when interrupts are enabled are equivalent to those produced in case (1), when interrupts are disabled.

\subsection{Security Argument -- \CFA engine}

Recall from Section~\ref{sec:model} that \adv takes advantage of vulnerabilities in the Normal World attempt forgery of \CFA reports. To attack the \CFA engine without interrupts, \adv must:  (i) modify or forge \CFlog; (ii) deactivate the \CFA engine by removing \app's instrumentation; or (iii) generate a \CFA report that is accepted by \vrf as authentic when it does not correspond to a timely (i.e., after issuance of the latest \chl) execution of \app on \prv.

\textbf{Forging \CFlog}. The trivial approach is directly modifying \CFlog in \prv's memory. However, \acro keeps \CFlog inside the Secure World, which is inaccessible to the Normal World. Another possibility is to call the trampoline function directly to add arbitrary entries to \CFlog. However, the code to perform these illegal trampoline calls cannot be injected in \app's program memory, since this region is made immutable by the Initialization Routine.
Therefore, \adv must jump from \app to malicious code that calls the trampoline (outside of \app) by corrupting the destination of an indirect branch. Since all indirect branches within \app are instrumented, this attempt would be appended to \CFlog and thus detectable by \vrf.

\textbf{Removing \app instrumentation}. Instrumentation integrity is ensured through $H(\app)$, included in $\report(\app)$. Once the \ISM Initialization Routine is called, {\bf before} \app execution, it computes $H(\app)$. \app's binary is unmodifiable thereafter (until \ISM finalization routine execution) due to the NS-MPU and \SAU protections enforced by \acro initial configurations. Thus, any attempt to modify \app must be made before the \ISM Initialization Routine call, resulting in an incorrect $H(\app^\adv)$ value that is detected by \vrf.

\textbf{Forging \CFA Report ($\report(\app)$)}. \adv may attempt to forge $\report(\app)$ to reflect the expected execution of \app when that execution did not happen. This requires finding $\sigma^{\adv}$ such that $Vrfy^{pk}(\sigma^{\adv}) = (H(\app)||H(\CFlog)||\chl)$. $sk$ is securely stored within the Secure world. Therefore, this forgery is computationally infeasible as long as $(Sig,Vrfy)$ is implemented using a cryptographically secure signature.

\subsection{Security Argument --  Interrupt Safety Module}

\ISM supports the \CFA engine by preserving the integrity of \CFA reports irrespective of interrupts. To bypass \ISM, \adv must (i) deactivate the \ISM Dispatcher;  (ii) replace an ISR address in the Non-Secure IVT; 
or (iii) corrupt \app's data in the Normal World stack.

\textbf{Deactivating \ISM Dispatcher}. \adv could attempt to remove the dispatcher integrity checks that must occur for each interrupt to safeguard \app's context. However, the Secure World controls the NVIC interrupt configuration, which cannot be altered by an \adv that compromises the Normal World. 

\textbf{Modifying Interrupt Control Flow}. When the Dispatcher redirects the execution to an untrusted ISR, it must eventually receive the control back to resume \app's context and execution. \adv could attempt to modify the ISR's return address and redirect the control flow, instead of returning to the Dispatcher or even add bogus entries to \CFlog by calling the trampoline function directly. \acro disables logging of new entries to \CFlog (through the LAC flag) and blocks \app's stack to the NS-World, pausing attestation until any ISRs return control to the Dispatcher. While attestation is paused, nothing can affect the \app's context. If the execution returns to the Dispatcher, it will recover the  \app's saved context and resume execution appropriately. Otherwise,  \acro will not produce a signed \CFA report, and \vrf would conclude that \app execution failed.

\textbf{Corrupting \app Stack.} When \acro Dispatcher is triggered  (due to an ISR) while \app executes in the Normal World, a hardware routine pushes the contents of several registers onto the Non-Secure stack. After the ISR execution ends, the Dispatcher loads a specific value into the program counter ($PC$) called EXEC\_RETURN. This particular value triggers an ``end of interrupt'' routine, where the hardware resumes the execution state to what it was before the interrupt was triggered, including its return address. Since the Non-Secure stack is accessible to the untrusted ISR, the saved return address could be corrupted, leading the Dispatcher to return to the wrong location. \adv could also change another context in the stack, such as saved register values and variables. By locking the Non-Secure stack belonging to \app and ensuring the integrity of the stack pointer, \acro guarantees that \app's context is appropriately resumed, including the ISR return address and the \app's registers and data memory values. 


\section{Evaluation}

We implement \acro's proof-of-concept prototype on a NUCLEO-L552ZE-Q~\cite{NUCLEOL552ZEQ} development board that is equipped with an STM32L552ZE MCU. This MCU is based on the ARM Cortex-M33 (v8) architecture and operates at a clock frequency of 110 MHz. The MCU does not feature a memory cache and supports Arm TrustZone-M technology. Our implementation includes  a python script to instrument the \app's assembly code, as required by the CFA engine. The trusted components, \ISM and \CFA engine, are implemented in \texttt{C} and run in the Secure World. An interface module that facilitates communication between the Secure and Normal World is also implemented in \texttt{C}. The CFA verification engine, executed by \vrf, is implemented in Python. In our experiments, we use a fixed \CFlog buffer of size 4096 Bytes, which is configurable to match the \app's requirements and available resources. Assuming a bare-metal system, our prototype features \acro running in the Secure World and \app running in the Normal World. In the Secure World, \acro operates in the Thread privileged mode, except for the Dispatcher, which runs in the Handler mode. Within the Normal World, \app runs in the Thread privileged mode, while its interrupts operate in the Handler mode. Section~\ref{sec:discussion} provides additional information on running \app or \acro using unprivileged mode within the context of a Rich/Trusted OS context.

We evaluate \acro prototype to (i) determine the time, memory and energy consumption overheads of each individual \acro module; (ii) determine what \acro's impact on real applications is; and (iii) test the approach against possible attacks.


\subsection{Runtime Overhead}
\label{sec:experiments:runtime overhead}
Four modules contribute to \acro overhead: Four modules contribute to \acro overhead: (1) \ISM Initialization Routine, to initialize attestation and hash the binary memory; (2) \ISM Dispatcher, which handles interrupts during the attested execution; (3) \ISM finalization routine, to finalize attestation and generate the \CFA report; and (4) \CFA engine, which appends new entries to \CFlog.

The runtime of each of the four modules is measured as follows:

\begin{compactitem}
  \item \ISM Initialization Routine: elapsed time measured from the instruction that calls the attestation (\ISM Initialization Routine) until the point where the first instruction of \app is reached.
  \item \ISM Finalization Routine:  elapsed time measured from \app's last instruction until the first instruction after finishing attestation. 
  \item \CFA engine: elapsed time measured from when the trampoline instruction is called until the execution of the first instruction after returning to the Normal World.
\end{compactitem}

\noindent The \ISM Dispatcher module is evaluated using two different metrics: 

\begin{compactitem}
  \item Interrupt Latency: time between interrupt triggering and execution of the corresponding ISR, including the trusted dispatcher execution time.
  \item Interrupt Backtrip Latency: time to resume \app execution after an ISR returns, including the dispatcher execution to assure \app's integrity and interrupt flag resets. 
\end{compactitem}


We measured the runtime of each \acro module $10^5$ times. The results are shown in Table~\ref{tab:mod elapse time}. The overhead due to the initialization and finalization routines is fixed as these modules are only executed once per attested execution. The primary source of overhead in these modules is the cryptographic operations (this cost can potentially be reduced significantly on devices with hardware-accelerated cryptographic instructions). The majority of the runtime in the \ISM Initialization Routine comes from the hash computation of $H(\app)$. In our measurements, we employed the blake2s\cite{Blake2scode} hash function and measured that a binary size of 1 KByte required 211864 CPU cycles, while the finalization routine took approximately 843816 CPU cycles for a \CFlog of size 4kBytes. The combined runtime of both modules increases linearly with the total size of \CFlog plus the binary (\app) size at a rate of 182103 CPU cycles per additional 1 KByte. This cost is associated to the hash and signature operations used to generate the \CFA report.

\begin{table}[]
  \centering
  \footnotesize
  \caption{Runtime and Energy Measurements of \acro}
  \vspace{-1em}
  \begin{tabular}{l|ccc}
  \hline
  \multirow{2}{*}{\textbf{Module}}                & \multicolumn{3}{c}{\textbf{\begin{tabular}[c]{@{}c@{}}CPU \\ Cycles\end{tabular}}} \\ \cline{2-4} 
                              & \textbf{Mean} & \textbf{Std} & \textbf{Max} \\ \hline
  \multicolumn{1}{c|}{ISM Initialization Routine} & 211865                     & 13                       &  211878                        \\
  ISM Finalization Routine    & 843816       & 13          & 843831         \\
  CFA engine                  & 491           & 8          & 502           \\ \hline
  \textbf{Without Dispatcher} &               &              &              \\
  Interrupt Latency           & 29          & 2          & 32           \\
  Interrupt Latency Backtrip  & 405        & 8          & 415           \\ \hline
  \textbf{With Dispatcher}    &               &              &              \\
  Interrupt Latency           & 103         & 5          & 110            \\
  Interrupt Latency Backtrip  & 452         & 8          & 463            \\ \hline
  \end{tabular}
  \vspace{-1.8em}
  \label{tab:mod elapse time}
  \end{table}

As depicted in Table~\ref{tab:mod elapse time}, each control flow logging event generated by the \CFA engine requires approximately 491 CPU cycles, primarily due to the context switches between the Normal and Secure World. The \CFA engine is triggered every time the execution reaches an instrumentation point (the beginning or end of each \CFG node). Consequently, the relative overhead (\%) is proportional to the ratio between branching and non-branching instructions. Instructions that do not branch do not require instrumentation and thus do not incur additional execution time. Thus, a greater number of non-branching instructions reduces the relative execution overhead. The worst case execution time (WCET) for the \CFA engine occurs when all of the instructions are branching instructions (a theoretical case that would not occur in real programs). This case would introduce an overhead of 491 CPU cycles for each instruction.

\begin{figure}[hbtp]
	\centering
   \vspace{-0.5em}
	\subfigure[Interrupt Latency]
	{\includegraphics[width=0.4\columnwidth]{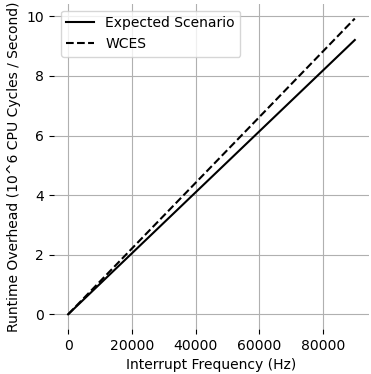}}
	\subfigure[Interrupt Backtrip Latency]
	{\includegraphics[width=0.4\columnwidth]{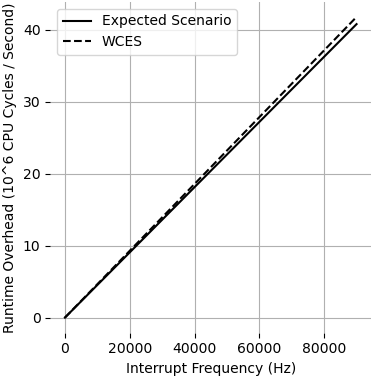}}
	\vspace{-0.6em}
	\caption{\acro Dispatcher runtime vs. interrupt frequency}
	\vspace{-1em}
	\label{fig:isc-flat overhead per freq} 
\end{figure}

To assess the efficiency of the \ISM Dispatcher, we measured the interrupt latency with and without it. Our results show that the interrupt latency introduced by the \ISM Dispatcher was approximately 103 CPU cycles, which results in 74 CPU cycles overhead compared to the baseline interrupt. However, we believe that this overhead is acceptable for most applications, considering the typical application latency requirements outlined in \cite{sbis}. Additionally, the interrupt backtrip latency was around 452 CPU cycles, which represents an increase of 46 CPU cycles over the baseline. These values are not affected by the ISR code size or  execution time, as they occur before and after the ISR execution. Therefore, the total runtime overhead introduced by \acro during an attested execution of \app depends on the frequency of interrupts. The expected overhead generated by the Dispatcher and the  WCET for different interrupt frequencies are shown in Figure~\ref{fig:isc-flat overhead per freq}.

\subsection{TCB Size}

Table 1 presents the number of lines of C/Assembly code (calculated using "cloc v1.90") implemented for each \acro module and the corresponding compiled binary size (optimization flag "-O0") that is incorporated into the TCB (including dependencies).
\acro implementation in the Secure World is written in a total of 542 lines of code (including an external Blake2s Library). As anticipated, the \CFA engine is a small portion of the \app binary due to its limited role of adding new entries to \CFlog. On the other hand, the \ISM Initialization and Finalization have larger code sizes, as they implement a range of operations, including cryptographic functions and register assignments. The Dispatcher, with an intermediate code size, performs fewer system configurations.
For a reference point, OAT~\cite{oat} reports 916 lines of C and assembly code.


\vspace{-.25cm}
\begin{table}[h]
    \centering
    \footnotesize
    \vspace{-1em}
    \caption{TCB size of \acro modules.}
    \vspace{-1em}
    \begin{tabular}{lcc}
    \hline
    \textbf{Module} &  \textbf{\begin{tabular}[c]{@{}c@{}}Lines\\ of Code\end{tabular}}  &\textbf{\begin{tabular}[c]{@{}c@{}}Binary Size\\ (Kb)\end{tabular}} \\ \hline
    \multicolumn{1}{c}{ISM Initialization Routine} & 242 & 6.31 \\
    ISM Finalization Routine                       & 73 & 1.78\\
    ISM Dispatcher                                 & 32 & 1.22\\
    CFA engine                                     & 20 & 0.38\\
    Blake2s Library                                & 177 & 3.34\\ \hline
    Total                                          & 562 & 13.03 \\ \hline
    
    \end{tabular}
      \label{tab : memory}
      \vspace{-1em}
    \end{table}

\subsection{Energy Consumption}
\label{sec:energy consumption}

\begin{figure}[h]
	\centering
	\subfigure[ISM Dispatcher]
	{\includegraphics[width=0.4\columnwidth]{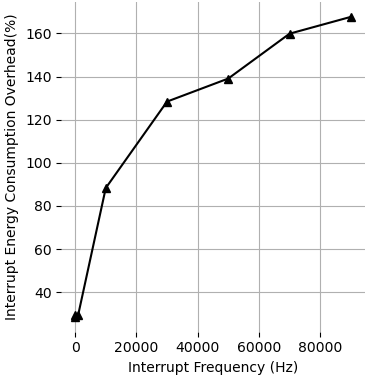}}
  \subfigure[Sleep Mode Consumption]
	{\includegraphics[width=0.4\columnwidth]{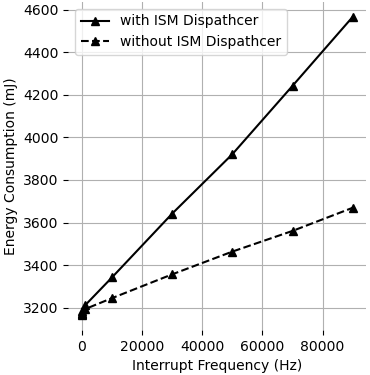}}
	\vspace{-0.6em}
	\caption{\ISM Dispatcher energy consumption.}
	\vspace{-2em}
	\label{fig:isc-flat energyoverhead} 
\end{figure}

To evaluate the energy consumption of \acro, we used a X-NUCLEO-LPM01A \cite{XNUCLEOLPM01A} evaluation board. This board powers the MCU with 3.3V, isolating it from the energy source of the other components on the MCU prototyping board. It measures the MCU's current usage with a 100kHz acquisition rate. To determine the energy consumption introduced by the Dispatcher at varying interrupt frequencies, we measured the MCU's current consumption during 100s of ``Sleep Mode'' with active interrupts. The ``Sleep Mode'' is deactivated when the interrupt handler executes and reactivated when it ends. We consider three configurations for the experiment: (i) no interrupts, (ii) with a single interrupt at different frequencies, and (iii) with the Dispatcher activated. The results of setups (ii) and (iii) are shown in Figure~\ref{fig:isc-flat energyoverhead}b. To isolate the Dispatcher overhead, we subtracted the consumption of setup (i) (without interrupts) from setups (ii) and (iii). The increment in (iii) compared to (ii) is shown in Figure~\ref{fig:isc-flat energyoverhead}a. Our results indicate that \acro introduces an overhead of 29\% at low-frequency interrupts, which increases to 167.7\% as the frequency approaches 90kHz.It is important to note that  the results in this subsection offer a micro-level perspective of interrupt energy consumption. The broader impact on the overall system will be assessed through case studies in Section~\ref{sec:cases}.



\subsection{Case Study Applications}\label{sec:cases}

In this section, we evaluate the performance of \acro interrupts on real embedded applications, aiming to demonstrate the effects of \acro overhead due to the \ISM module. \acro is the first interrupt-safe CFA, thus there are no approaches that achieve the same functionality. Therefore, we compare it to regular \CFA to account for the additional cost of supporting interrupts safely. Our evaluation considers the overhead introduced by the \CFA engine (required by any TEE-based \CFA method, irrespective of interrupt-safety) over the baseline application without any \acro modules. Most importantly, it isolates the overhead due to the \ISM module (our main contribution) over the baseline \CFA engine. Finally, we also experiment with \acro defenses against different interrupt attack modes. 
\begin{table*}[]
    \centering
    \footnotesize
    \caption{Overhead Generated by \acro modules over the baseline.}
    \vspace{-1em}
    \begin{tabular}{ccccc|cccc|cccc}
    \hline
    \multicolumn{1}{l}{}        & \multicolumn{4}{|c|}{E1 -- Syringe Pump}           & \multicolumn{4}{c|}{E2 -- UltrasonicRanger} & \multicolumn{4}{c}{E3 -- Fire Sensor}  \\ \hline
     &
      \multicolumn{2}{|c|}{\begin{tabular}[c]{@{}c@{}}CFA\\ Engine\end{tabular}} &
      \multicolumn{2}{c|}{\begin{tabular}[c]{@{}c@{}}CFA Engine\\ + ISM\end{tabular}} &
      \multicolumn{2}{c|}{\begin{tabular}[c]{@{}c@{}}CFA\\ Engine\end{tabular}} &
      \multicolumn{2}{c|}{\begin{tabular}[c]{@{}c@{}}CFA Engine\\ + ISM\end{tabular}} &
      \multicolumn{2}{c|}{\begin{tabular}[c]{@{}c@{}}CFA\\ Engine\end{tabular}} &
      \multicolumn{2}{c}{\begin{tabular}[c]{@{}c@{}}CFA Engine\\ + ISM\end{tabular}} \\ \hline
    \multicolumn{1}{c|}{\begin{tabular}[c]{@{}c@{}}Interrupt\\ Frequency\end{tabular}} &
      \begin{tabular}[c]{@{}c@{}}Runtime\end{tabular} &
      \multicolumn{1}{c|}{\begin{tabular}[c]{@{}c@{}}Energy \end{tabular}} &
      \begin{tabular}[c]{@{}c@{}}Runtime\end{tabular} &
      \begin{tabular}[c]{@{}c@{}}Energy \end{tabular} &
      \begin{tabular}[c]{@{}c@{}}Runtime\end{tabular} &
      \multicolumn{1}{c|}{\begin{tabular}[c]{@{}c@{}}Energy \end{tabular}} &
      \begin{tabular}[c]{@{}c@{}}Runtime\end{tabular} &
      \begin{tabular}[c]{@{}c@{}}Energy \end{tabular} &
      \begin{tabular}[c]{@{}c@{}}Runtime\end{tabular} &
      \multicolumn{1}{c|}{\begin{tabular}[c]{@{}c@{}}Energy \end{tabular}} &
      \begin{tabular}[c]{@{}c@{}}Runtime\end{tabular} &
      \begin{tabular}[c]{@{}c@{}}Energy \end{tabular} \\ \hline
    \multicolumn{1}{c|}{100Hz}  & 19.3\% & \multicolumn{1}{c|}{9.1\%} & 19.3\% & 9.1\% & 4.5\% & \multicolumn{1}{c|}{2.4\%} & 4.5\% & 2.4\% & 10.2\% & \multicolumn{1}{c|}{4.9\%} & 10.2\% & 4.9\%\\
    \multicolumn{1}{c|}{1kHz}   & 19.3\% & \multicolumn{1}{c|}{9.1\%} & 19.3\% & 9.1\% & 4.5\% & \multicolumn{1}{c|}{2.4\%} & 4.5\% & 2.4\% & 10.2\% & \multicolumn{1}{c|}{4.9\%} & 10.3\% &   4.9\%\\
    \multicolumn{1}{c|}{10kHz}  & 19.5\% & \multicolumn{1}{c|}{9.2\%} & 20.9\% & 10.1\% & 4.6\% & \multicolumn{1}{c|}{2.4\%} & 5.7\% & 2.6\%  & 10.5\% & \multicolumn{1}{c|}{4.9\%} & 10.9\% &  5.1\%\\
    \multicolumn{1}{c|}{30kHz}  & 21.3\% & \multicolumn{1}{c|}{9.7\%} & 25.2\% & 12.9\% & 6.5\% & \multicolumn{1}{c|}{3.6\%} & 8.4\% & 3.7\% & 11.7\% & \multicolumn{1}{c|}{5.4\%} & 13.1\% &     6.7\%\\
    \multicolumn{1}{c|}{50kHz}  & 24.8\% & \multicolumn{1}{c|}{11.5\%} & 29.6\% & 14.7\% & 10.3\% & \multicolumn{1}{c|}{5.1\%} & 16.1\% & 6.4\% & 13.0\% & \multicolumn{1}{c|}{6.1\%} & 18.8\% &  8.6\%\\
    \multicolumn{1}{c|}{70kHz}  & 28.7\% & \multicolumn{1}{c|}{14.0\%} & 35.1\% & 17.5\% & 13.7\% & \multicolumn{1}{c|}{6.3\%} & 18.8\% & 9.0\% & 15.8\% & \multicolumn{1}{c|}{7.3\%} & 21.6\% &  9.9\%\\
    \hline
    \end{tabular}
    \vspace{-1.5em}
    \label{tab:case overhead}
    \end{table*}

In order to assess \acro in real-world scenarios, we customized three open-source applications commonly utilized for evaluating previous research on CFA. These include: \textbf{\textit{[E1]}} a Syringe Pump~\cite{syringepump} that manages a fluid injection medical device,  \textbf{\textit{[E2]}} an Ultrasonic Ranger~\cite{UltrasonicRanger} used to measure the proximity of obstacles for parking assistance applications, and  \textbf{\textit{[E3]}} a Fire Sensor~\cite{tempandhumi} implemented based on a combination of temperature and humidity measurements.

 In our experiments, we incorporated a timer interrupt that operates independently from the application. This interrupt retrieves the sensor readings at a frequency of $f$, and stores them in a buffer. For each frequency of $f$ = {100Hz, 1kHz, 10kHz, 30kHz, 50kHz, 70kHz}, we repeated the experiment 100 times. For each test, we evaluated the performance of the three applications, measuring their runtime and energy consumption under different conditions: without any \CFA (baseline), with the \CFA engine only, and with both the \ISM and \CFA engines activated. Afterward, we calculated the overall time and energy consumption overhead incurred by the modules compared to the baseline experiment. Table \ref{tab:case overhead} presents the results of our experiments. Consult \cite{regehr2005preventing} as a reference on the worst-case interrupt frequency across various real-life applications.

\textbf{Runtime Overhead}.
The \CFA engine incurs non-negligible runtime overhead primarily due to the frequent invocations of the Secure World (twice per execution of a \CFG node). The overhead observed for the \CFA engine is consistent with prior research on TEE-based \CFA. On the other hand, the ISM overhead added atop the CFA engine is negligible when utilizing low interrupt frequencies (ranging from 10 to 1kHz). Our results show an overhead of $< 0.1\%$. The overhead becomes noticeable when the frequency reaches 10kHz, where it slowly rises from a range of 0.4\% ($E3$) to 1.4\% ($E1$) and reaches a maximum overhead between 5.1 \% ($E2$) to 6.4\% ($E1$) at around 70kHz. In sum, these results indicate that \acro is well-suited for systems with low-frequency interrupt requirements (below 10kHz) and incurs modest overhead at higher frequencies.

\textbf{Energy Consumption Overhead}.
The energy usage of each application is evaluated using the same configuration and board as specified in Section~\ref{sec:energy consumption}. As shown in Table~\ref{tab:case overhead}, the relative energy consumption overhead is roughly half of the runtime overhead. These findings demonstrate the suitability of \acro for systems with low frequency (below 10kHz) and low-power devices while still allowing an acceptable energy expenditure for a diverse range of applications on high interrupt frequency settings.

\textbf{Security Tests}. 
We design malicious interrupts to launch control flow hijacks, including the examples presented in Section~\ref{fig:problem_setting_examples}, within the Syringe Pump execution and validate the effectiveness of \acro by analyzing its measurements. Specifically, we analyzed the following attack vectors: 

{\it -- Attacking the return address to redirect interruptions:} our malicious interrupts attempt to redirect the interrupts by changing the IRS return address and the Dispatcher address. When it tries to modify the Dispatcher return addresses, a fault exception is generated due to illegal MPU area access. By changing the return address of the ISR, we could redirect the execution. However, \adv is unable to resume \app execution. Therefore, it can not generate a signed \CFA report and it runs in a different context, unrelated to \app. 

{\it -- Deactivating \acro configurations:} This case attempts to misconfigure the MPU and the NS-IVT to bypass \acro. This case always generates a fault exception due to an illegal access to a Secure World region and illegal access to MPU-protected region, respectively.

Unfortunately, there are no public/widely used benchmarks for this kind of attack yet. Thus we were unable to perform an independent analysis and had to implement our own attack vectors.
Nonetheless, we hope that these attack vectors serve as a proof of concept to show \acro security in practice.
They are also included in our public release of \acro's implementation~\cite{repo}.
\vspace{-.2em}
\section{Discussion \& Limitations}
\label{sec:discussion}
\textbf{Memory and Performance Optimization}. \CFA approaches may generate large \CFlog-s, depending on \app's behavior. Instrumenting every node in \app can also lead to many TrustZone calls and added runtime overhead. The \CFA engine in \acro is generic and replaceable. Therefore, new techniques to optimize \CFlog construction can be merged into \acro without interfering with the interrupt-safety provided by \ISM. In this work, we do not focus on optimizing the \CFA engine. We refer the reader to~\cite{oat} for a comprehensive treatment of optimization issues.

\textbf{Non-Control Data Attacks}. \CFA detects and reports all control flow deviations.  Nonetheless, non-control data-only attacks \cite{269251,7546545} can exploit specific vulnerabilities to corrupt application data without modifying \app's control flow path. While these attacks require less likely vulnerabilities (e.g., ``write anywhere'' bugs), they are still possible. To detect such cases, future work should consider interrupt-safe attestation of both control and data flows.

\textbf{Coexistence with RTOS}. It is worth noting that RTOS-s rely on interrupts to control task scheduling. Thus, by isolating \app against interrupts, \acro can isolate \app from other concurrent applications (or the RTOS) on the same device. The dispatcher would consider interrupts from the RTOS as untrusted ISRs and preserve \app state across them. Nonetheless, to support \acro implementation along with an RTOS, certain compatibility challenges must be addressed. The simplest version of \acro precludes the (Normal-World) OS from reconfiguring NS-MPU during the attested process execution. In this case, the (Normal-World) OS should request NS-MPU re-configuration to the Secure-World. Furthermore, the initialization interface must account for OS-specific process management policies.

\textbf{Coexistence with Trusted OS}.
\acro's proof-of-concept prototype currently runs on bare-metal. However, it could also be implemented within a Secure-World firmware or operating system, such as TF-M~\cite{TF-M}. If implemented as a software module, \acro would become part of the trusted firmware's or operating system's TCB, increasing its size accordingly. Alternatively, running \acro as a trusted service or application in an unprivileged Thread mode would avoid increasing the TCB size but could potentially result in increased runtime overhead due to limited access to resource for the Dispatcher.

\vspace{-1em}
\section{Conclusion}

This paper characterizes the conflict between real-time application needs and existing TEE-based \CFA methods. We also demonstrate interrupt-based attacks on typical \CFA designs. Motivated by this problem, we propose \acro: a \tee-based interrupt-safe \CFA scheme. \acro protects the integrity of interrupted attested programs against vulnerable/malicious interrupts. It tracks and controls interrupt access within the \tee's secure world to ensure that interrupt service routines cannot tamper with the integrity of \CFA reports. We implement and evaluate \acro and make its implementation publicly available.

\section*{Acknowledgments}

The authors thank the anonymous reviewers and designated shepherd for their guidance and feedback. This work was supported by the National Science Foundation (Award \#2245531) as well as a Meta Research Award (2022 Towards Trustworthy Products in AR, VR, and Smart Devices RFP).

\balance

\bstctlcite{IEEEexample:BSTcontrol}
\bibliographystyle{IEEEtranS}
\bibliography{references}
\end{document}